\documentclass{ws-rv9x6}
\usepackage{subfigure}   
\usepackage{ws-rv-thm}   
\usepackage{ws-rv-van}   

\makeindex

\newcommand{\ecm}{\mbox{$e$\,cm}}
\newcommand{\efm}{\mbox{$e$\,fm}}
\newcommand{\half}{\mbox{$\frac{1}{2}$}}
\newcommand{\Hel}{\mbox{$^3\text{He}$}}
\newcommand{\Tri}{\mbox{$^3$H\phantom{e}}}
\newcommand{\he}{{}^{3}\text{He}}
\newcommand{\trit}{{}^{3}\text{H}}
\newcommand{\NME}{\mathcal{F}}

\begin{document}

\chapter[Permanent EDMs of 1-, 2-, and 3-Nucleon Systems]{Permanent Electric Dipole Moments of Single-,
Two-, and Three-Nucleon  Systems}\label{sec: title}

\author[A. Wirzba, J. Bsaisou and A.Nogga]{Andreas Wirzba$^{1}$\footnote{a.wirzba@fz-juelich.de}$^{,2}$, Jan Bsaisou$^1$ and
Andreas Nogga$^{1}$\footnote{a.nogga@fz-juelich.de (on  leave of absence from the Forschungszentrum J\"ulich)}$^{,3}$}

\address{$^1$ IKP-3 and IAS-4,
               Forschungszentrum J\"ulich,\\ 
               D-52425 J\"ulich, Germany\\
               $^2$ Kavli Institute for Theoretical Physics,
               University of California,\\ 
               Santa Barbara,  CA 93106-4030, USA\\
               $^3$ Institute of Nuclear and Particle Physics\\ and Department of Physics and Astronomy, Ohio University,\\
               Athens, OH 45701, USA}

\begin{abstract}
A nonzero  electric dipole moment  (EDM) of the neutron, proton, deuteron or helion,  in fact, of any finite system necessarily involves the breaking of a symmetry,  either by the presence of external fields (i.e.\ electric fields  leading to the case of {\em induced} EDMs) or explicitly by the breaking of the discrete parity and time-reflection symmetries in the
case of {\em permanent} EDMs.
We discuss two theorems describing these phenomena and report about the cosmological motivation for an existence of $CP$ breaking beyond what is generated by
the Kobayashi-Maskawa mechanism in the Standard Model and what this might imply for the permanent electric
dipole moments of the nucleon and light nuclei by estimating a  window of opportunity 
for physics beyond what is currently known.
Recent -- and in the case of the deuteron even unpublished --  results for the relevant 
matrix elements
of nuclear EDM operators are presented and the relevance for disentangling underlying New Physics sources are discussed.
\end{abstract}

\body


\section{The Problem with Permanent Electric Dipole Moments}\label{sec: intro}
Gerry Brown was always interested in
 {\em magnetic} dipole moments of baryons and nuclei, and especially in confronting the
theoretical predictions of quark, chiral bag and Skyrme models with the experimental results. 
But to our knowledge (compare also with Ref.~\refcite{Wirzba:2014mka}), he never worked
on {\em electric} dipole moments. This choice  definitely turned out to be wise in his case,
since during his lifetime  -- and
even until the time of writing --  all
measurements of the electric dipole moment of any (sub-)atomic particle have been compatible with zero -- only more and more
restrictive upper bounds have
been established since the first experiment in the fifties of the last century by Smith, Purcell, and 
Ramsey\cite{Ramsey} for the neutron EDM.

\subsection{The subtle character of EDMs of subatomic particles}
Why are  permanent electric dipole moments (EDMs), which in classical electrodynamics just correspond
to  (the integrals over) the spatial three-vectors of displaced charges (or in general charge densities), 
much
more subtle in the case of subatomic particles or, generically, in the realm of Quantum Mechanics?
The reason is that their existence is tied to the breaking of the discrete symmetries of parity ($P$) conservation and time-reflection ($T$) invariance, such that   they are  intrinsically small.\footnote{The breaking of $T$ implies $CP$ violation (in terms of the charge conjugation $C$ symmetry) if $CPT$ is conserved.} 
 In fact, the order of magnitude of the nucleon EDM $(d_{\text{N}})$  can be estimated as follows:\cite{Lamoreaux}
\begin{Romanlist}[II] 
\item  The  starting scale is given by 
 the $CP$ and $P$ conserving (magnetic) moment of the nucleon, which is of the order of 
 the nuclear magneton 
 \begin{equation}
   \mu_{\text{N}} = {e}/({2 m_{\text{p}}}) \sim 10^{-14} \,\ecm\,,
  \end{equation} 
  where $e>0$ is the unit of electric charge and $m_{\text{p}}$ the proton mass.
\item Furthermore, as we will discuss below, a nonzero permanent EDM requires  $P$ and $CP$  violation.
   The cost  of $P$ violation can  empirically be estimated in terms of
   Fermi's constant  $G_{\text{F}} \approx 1.166\cdot 10^{-5} \,\text{GeV}^{-2}$ times the square of the axial
  decay constant of the pion, $F_\pi \approx 92.2\,\text{MeV}$, the order-parameter of the spontaneous breaking of chiral symmetry of Quantum Chromodynamics (QCD)
 at low energies\cite{pdg:2016}. The dimensionless product scales therefore as 
 $
      G_{\text{F}} \cdot F_\pi^2  \sim 10^{-7}$.
\item
  The cost related to  the {\em additional} $CP$ violation follows from, e.g., the ratio of the amplitude moduli
 of $K_{\text{L}}^0$ to $K_{\text{S}}^0$ decays into
 two pions\cite{pdg:2016}:
 \begin{equation}
    |\eta_{+-}| = \frac{| \mathcal{A}(K_{\text{L}}^0 \to\pi^+\pi^-)|}{|\mathcal{A}(K_{\text{S}}^0 \to \pi^+ \pi^-)|} = (2.232\pm 0.011)\cdot 10^{-3}
        \,.
 \end{equation}
\end{Romanlist}
In summary, the modulus of the EDM of the nucleon cannot be larger than 
 \begin{equation}
   | d_{\text{N}} | \sim 10^{-3} \times 10^{-7}\times \mu_{\text{N}} \sim 10^{-24}\, \ecm\,,
    \label{nEDM-estimate}
\end{equation} 
which is ten orders of magnitude smaller than the corresponding magnetic dipole moments,
 without getting into conflict with known physics  --- on top of the EDM measurements themselves which are nowadays even more restrictive (see below).

 In the Standard Model  the sole source for $CP$ violation, if the QCD $\theta$ term is assumed to be absent, is the
 Kobayashi-Maskawa\cite{Kobayashi:1973fv} (KM) mechanism, which, however,
 only generates a  nonzero $CP$ violating phase if  the Cabbibo-Kobayashi-Maskawa (CKM) quark-mixing matrix involves at least three quark generations.
 This KM-generated $CP$ violation is therefore  flavor-violating, while  the EDMs
 are, by nature, flavor-diagonal. This means that the SM (without the QCD $\theta$ term) is ``punished'' by the
 additional cost of a further
 factor $G_{\text{F}} F_\pi^2 \sim 10^{-7}$ to undo the flavor violation.  In summary, the SM prediction for the nucleon EDM, based on the KM mechanism, 
 is therefore much smaller than \eref{nEDM-estimate}, namely
 \begin{equation}
   | d_{\text{N}}^{\text{SM}} | \sim 10^{-7} \times 10^{-24}\, \ecm \sim 10^{-31}\,\ecm\,.
    \label{nEDM-estimate-SM}
\end{equation} 
This  result agrees in magnitude with the three-loop estimates of Refs.~\refcite{Khriplovich:1985jr,Czarnecki:1997bu} 
and  also
with the two-loop calculations of Refs.~\refcite{Gavela:1981sk,Khriplovich:1981ca} (see also Ref.~\refcite{Eeg:1983mt})
which include both a strong {\em penguin}  short-range diagram and a long-range pion loop.\footnote{Note that one-loop contributions
to EDMs resulting from the KM mechanism of the SM have to vanish, since the  $CP$-violating KM matrix element at the first loop vertex is
canceled by its Hermitian conjugate at the other side.} 
 Even recent {\em loop-less} calculations\cite{Mannel:2012hb}
 involving propagators of  
charm-flavored sea-quarks   give
a result of the same order.\footnote{The EDM  of the electron is even further suppressed by a factor $10^{-7}$ in the SM, 
i.e.\ 
 $|d_e^{\text{SM}}| \sim 10^{-38}\,\ecm$, which follows from a further  weak-interaction insertion and one additional
 quark/gluon loop\cite{Pospelov:1991zt}. The SM prediction for the muon is slightly larger, namely $|d_\mu^{\text{SM}}| \sim
 10^{-35}\,\ecm$, because of the lepton mass ratio $m_\mu/m_e \sim 200$.}
 
 From  the above estimates one can infer that an EDM of the nucleon measured in the
 window
  \begin{equation}
      10^{-24} \ecm  > |d_{\text{N}}| \gtrsim 10^{-30} \ecm 
      \label{window}
 \end{equation}
could be a clear signal for new physics beyond the KM mechanism of the Standard Model: either {\em strong $CP$ violation}
by  a sufficiently large
QCD $\theta$ term or genuinely new physics, as, e.g., supersymmetric (SUSY) models, multi-Higgs models, left-right-symmetric models etc.

\subsection{Two theorems for the existence of EDMs}
This brings us back to the original question: Why do nonzero electric dipole moments of finite quantum systems necessarily require the breaking of some symmetry?
The above statement can be interpreted as a special case of the following theorem which, e.g., is well-known to apply
for the case of the chiral symmetry breaking for lattice QCD (see e.g. 
Refs.~\refcite{Leutwyler:1987ak,Gasser:1987zq,Gross:1996}):\footnote{In a finite lattice scenario, even when the lattice size becomes
larger and larger,
a non-zero value of a quark condensate can only be measured  if the mass of the pertinent current quark differs
from zero corresponding to an {\em explicit}
breaking of chiral symmetry.}
\begin{theorem}
In {\em any  finite} quantum system  in the {\em absence} of any explicitly broken symmetry there cannot exist a spontaneously broken
ground state.
\end{theorem}
The condition of {\em finiteness} applies to both the spatial extent of the system and to the height of the pertinent quantum ``walls''. Therefore the  tunneling probability from {\em any} broken-symmetry state to any {\em alternative} one  is non-vanishing. 
This opens up the way for  finite systems to form one totally symmetrized state from all the broken-symmetry alternatives which is then the real ground state of the system, while the same tunneling amplitudes induce non-vanishing gaps to suitable  antisymmetric combinations of these states which are therefore excited states and non-degenerate to the ground state.

The question might arise why this does not apply to magnetic moments which have 
known and very well-tabulated\cite{pdg:2016}  nonzero values for the cases of the electron, muon, proton, neutron, other baryons, nuclei etc. 
In fact, the theorem applies but the solution is trivial. The non-zero value of the total angular momentum (i.e. the spin in the case of subatomic particles) suffices to induce the (rotational) symmetry breaking since 
it defines an axis  (in the laboratory frame) for the projection of 
the magnetic moment,\footnote{More precisely, for any non-zero spin and any direction in space
it is always possible to find  an eigenstate of the spin operator with {\em non-vanishing} 
projection (quantum number).} 
which shares the same axial-vector properties as the spin. 
The appearance of a nonzero magnetic moment for a particle without any spin or angular momentum is forbidden and would indeed come as a surprise. 

So what is the difference to
the case of an electric dipole moment which has the operational definition of the displacement vector of the
charges? Why is the existence of a nonzero spin of a particle or finite quantum system not enough to induce the
necessary symmetry breaking? The difference is that spins and, in general, angular momenta $\vec J$ are 
 {\em axial} vectors (as the magnetic fields $\vec B$) while the electric dipole moment  $\vec d $ is a {\em polar} 
 vector:
\begin{equation}
\begin{array}{ccc}
     \vec J & \stackrel{P}\longleftrightarrow & \phantom{-}\vec J \,,\\
     \vec J & \stackrel{T}\longleftrightarrow &-\vec J\,,
     \end{array} \qquad 
     \begin{array}{ccc}\vec d &\stackrel{P}\longleftrightarrow & -\vec d \,,\\
     \vec d & \stackrel{T}\longleftrightarrow & \phantom{-}\vec d\,.
     \end{array}
\end{equation}
 Without the explicit
breaking of the discrete $T$ (time-reflection) and $P$ (parity) symmetries, the presence of a non-zero spin
or angular momentum would not be enough  to define the direction for the projection of the EDM vector, since
the sign of this  projection would be reversed under the above mentioned discrete symmetries if they are conserved.
However, in the rest-frame of any subatomic particle
with non-vanishing mass there are simply not  any other vectors than the spin and total angular momentum.
 So it should be clear that
at least in these cases there is a need for extra symmetry breaking if these particles are to carry a non-vanishing
permanent EDM. The word {\em permanent} is important here, since the realization of an {\em induced} EDM
is of course possible -- without the breaking of $T$ and $P$  -- in the presence of a non-vanishing electric field $\vec E$ which has the properties of a polar vector as the EDM.
Note that, interpreted in this way, our theorem still holds also for the case of {\em induced} EDMs: the presence
of an electric field, which by nature breaks the (rotational) symmetry of the system, is the stated precondition.

But we know that macroscopic  and mesoscopic devices (capacitors,  batteries, etc.) and even certain molecules (H$_2$O or NH$_3$) obviously
can have sizable dipole moments which correspond to their spatial extent times the involved charges. 
Well, most of these systems break a symmetry classically. There is e.g.
a spatial vector pointing from one plate of a capacitor to the other one or from one nucleus in a diatomic polar molecule to the other one (which differs in charge and/or mass). But if these systems are interpreted
quantum mechanically,  as axially symmetric rods
or as (a)symmetric tops,  one should keep in mind the difference 
between body--fixed directions and lab-frame--fixed ones. If a polar symmetry still applies,
the projection on a stationary state
of fixed angular momentum in the lab-fixed frame suffices to average out the body-fixed (classical or intrinsic) EDM to
a vanishing  expectation value.\cite{Lamoreaux,Auerbach:1996zd,Spevak:1996tu}.
In the (a)symmetric top scenario, the tunneling amplitude from the
state pointing in one direction of the lab-frame (spin) axis to the one projected  onto the opposite direction would
be small, but nonzero,\cite{Lamoreaux} such that even then the theorem applies in principle. 
A nonzero EDM is measured in practice in  the latter cases, since the applied electric fields
might be small but non-vanishing, such that  the measured EDM has the character of an induced 
EDM.
 Alternatively, the pertinent temperature of the system is nonzero or the system, because it might be unstable,  it might have finite
 level widths or the measuring time might be not long enough to resolve the single levels, especially if  the tunneling
 gaps were tiny.\cite{Greene:2000,W.Li:2011} In this way the resulting de-facto
degeneracy between the parity-even ground state and excited parity-odd states together with the direction  defined  either by
the non-vanishing electric field or solely by the total-spin direction would be sufficient 
to define the orientation for the resulting explicit symmetry breaking. If however the system were cooled down to such small temperatures and sufficiently 
shielded against  electric fields and observed for a long enough time
that the de-facto degeneracy would be lifted,  a non-vanishing EDM 
could not  be measured even in these  cases. 
Only a truly infinite system would escape the consequences of the theorem.

Let us summarize what was stated above by the following theorem  which describes the existence of  
  permanent EDMs:\cite{Greene:2000,Bernreuther:2012,Wirzba:2014mka}
 \begin{theorem}
  \noindent Any non-vanishing coefficient $d$ 
   in the relation of the expectation values 
  $ \langle j^P | \vec d|  j^P \rangle = d \langle j^P| \vec J | j^P \rangle$ 
  of  the  electric dipole moment operator $\vec d \equiv \int \vec r \rho(\vec r) d^3 r$ (where $\rho(\vec r)$ is the
  charge density)
   and the total angular momentum $\vec J$ expressed in terms of 
   a stationary state $|j^P\rangle$ of a particle  
   with  at least one nonzero generalized `charge',
    nonzero mass, 
   nonzero total angular momentum 
   $j$ 
   and specified parity $P$,
   such that $\langle j^P| \vec J | j^P \rangle\neq 0$  in general,
   and no other energy degeneracy
  than its rotational one 
   is  a signal
  of $P$ and $T$ violation\footnote{Without the violation of $P$, 
    $\langle j^P | \vec d|  j^P \rangle$ would just vanish since 
${\cal P} |j^p\rangle = (-1)^P |j^P\rangle$ 
and ${\cal P} \vec d\, {\cal P} = - \vec d$, 
where ${\cal P}$  is the parity operator
which has the property 
${\cal P}^2 = 1$.}
 and, because of the $CPT$ theorem, of flavor-diagonal $CP$ 
  violation.  
  \end{theorem}
The above particle can be an {\em `elementary'} particle as a quark, charged lepton, $W^\pm$ boson, Dirac neutrino, etc., or  a {\em `composite'} particle as a neutron, proton, nucleus, atom, molecule or even a solid body, as long as it meets the requirements stated in the above theorem.
Namely,
it is important  that 
\begin{romanlist}[iii]
\item the particle or system should carry a non-vanishing angular momentum to define an axis (excluding therefore
scalar and  pseudoscalar particles),
\item
it should not be self-conjugate\footnote{Examples for {\em self-conjugate} particles with spin are  Majorana neutrinos or the $\omega$, $\rho^0$ or $\phi^0$ mesons, but not their SU(3) partners $K^{\ast}$  which carry strangeness
quantum numbers.} 
 in order  to prevent that the charge-conjugation property of the particle does not even allow a unique orientation in its body-fixed frame, 
\item
it should be in a stationary state (i.e.\ the observation time and, in case it is a resonance, its lifetime should be so large
that the pertinent energy level including its width  has not any overlap with  the levels of other states of opposite parity),
\item 
and 
that there should
be no degeneracy (except of  the states which only differ in their  {\em magnetic} quantum numbers and which have the same parity of
course). 
Otherwise the ground state of even parity could mix with a state of opposite parity and the directional
information coming from the spin would suffice to define the quantization axis for the EDM without the need of explicit $P$ and $T$ breaking. 
\end{romanlist}
However, it is well known and especially applies to the case of molecular systems with closely 
spaced rotational levels 
or atoms with a sizable octupole moment that the near-degeneracy of the two opposite-parity
levels might produce sizable enhancement factors for  $P$ and $T$ violating 
quantities, see e.g.\ Refs.~\refcite{Auerbach:1996zd,Spevak:1996tu}. A similar mechanism is  at work in the case of the {\em induced} EDMs of water or
ammonia molecules, see e.g.\ Ref.~\refcite{Bernreuther:1990jx}: In a simplified picture there is
a pair of nearly degenerate states of opposite parity $|\pm\rangle$ (where $|+\rangle$ is the ground state) 
with energy levels which rearrange according to
\begin{equation}
E_{2,1} = \half(E_- + E_+) \pm \sqrt{{\textstyle\frac{1}{4}}(E_--E_+)^2 +  (e\langle \vec r \,\rangle \cdot \vec E\,)^2}
\end{equation}
when exposed to an electric field $\vec E$.  Here
$\langle \vec r\,\rangle$ is the {\em transition} (not a diagonal!) matrix element of the charge displacement 
vector $\vec r$ between the states $|+\rangle$ and $|-\rangle$. For a sufficiently large $\vec E$ field, the second term in
the square root dominates and there will be an approximately linear behavior of the levels,
$
E_{2,1} =  \half(E_- + E_+) \pm |e \langle \vec r\, \rangle \cdot \vec E\, |$,
 which mimics a linear Stark effect -- but note  the appearance of an absolute value. For a weak enough $\vec E$-field and a sufficiently low temperature we would
find instead the following behavior (quadratic Stark effect):
\begin{equation}
E_{2,1} = E_\mp \pm \frac{(e \langle \vec r\, \rangle \cdot \vec E\,)^2}{E_- -E_+} + \cdots\,.
\end{equation}
Thus the  molecule  has always  an  {\em induced} EDM 
which can be enhanced by the small energy difference between  the states of opposite 
parity.\footnote{The case of a two-level system with a magnetic-moment  interaction in  the presence 
of a magnetic field 
is totally different, since there is  always a linear contribution $\langle\pm |\vec \mu\,|\pm \rangle \cdot \vec B$, no matter how weak the  $\vec B$ field might be, because these expectation values are  diagonal.}

\section[Motivation]{Motivation for EDMs}\label{sec: intro2}
Why should we be interested in measuring {\em permanent} EDMs? One reason is of course the
window of opportunity which the tiny $CP$-violating Kobayashi-Maskawa mechanism of the SM opens  for the
search of {\em New Physics}, see relation (\ref{window}). The other reason is 
the $CP$ violation by itself.
Independently of how much matter surplus might have originally been created in the Big Bang,
after the  inflation epoch the primordial  baryon--antibaryon (density) asymmetry should have  
been leveled out to an extremely high precision.
 However, about 380000 years later, when the temperature of the Universe had  sufficiently decreased such that 
 hydrogen atoms became stable against the radiation pressure and therefore the corresponding photons could not couple any longer to an electron-proton plasma,
 the ratio of this asymmetry to the  photon density $n_\gamma$ had the following measured
value:
\begin{equation}
 \frac{n_B - n_{\bar B} }{n_\gamma}\Big |_{\text{CMB}} = (6.05 \pm 0.07)\cdot 10^{-10} \,.
\end{equation}
This result was derived from the cosmic microwave background (CMB) measurements by the 
COBE, WMAP and Planck satellite missions\cite{Ade:2015xua}, 
while  the prediction of the SMs of particle physics and cosmology
 is more than seven orders of magnitude less, see e.g.\ Ref.~\refcite{Bernreuther:2002uj}.
 
 In fact, $CP$ violation is one of the three conditions for the dynamical generation of the baryon--antibaryon asymmetry 
 during the evolution of the universe as formulated by Andrey Sakharov in 1967\cite{sakharov}.
 These conditions can be paraphrased as follows:
 \begin{romanlist}[(ii)]
\item There has to exist a mechanism for the generation of baryon charge $B$ in order 
to depart from the initial value $B=0$ (after 
inflation).
\item Both $C$ and $CP$ have to be violated such  that  the production mechanisms and rates of $B$  
can be distinguished from the  ones of  $\bar B$ (even then  the pertinent helicities are summed).
\item  Either $CPT$ has to be broken as well\footnote{This in turn would imply the violation of Lorentz-invariance 
or locality or hermiticity.} or the dynamical generation had to take place during a stage of  non-equi\-librium
(i.e.\ a sufficiently strong first order phase transition) to discriminate, in the average, the $B$ production reaction from its back reaction and to escape  from the fact
that  $\langle B \rangle =0$ holds on the average if $CPT$ symmetry holds, i.e. from the time-independence in the equilibrium phase.
\end{romanlist}

While baryon plus lepton number ($B+L$) violation can be accommodated by
the  Standard Model in an early stage of the evolution via sphalerons,\cite{Kuzmin:1985mm}
the SM 
cannot satisfy the other two conditions: 
\begin{romanlist}[iii]
\item the
$CP$ breaking by  the Kobayashi--Maskawa (KM) mechanism\cite{Kobayashi:1973fv}
 is far too small; 
 even a $\bar\theta$ angle\footnote{The QCD parameter $\bar\theta$ is the sum of the original angle of
 the QCD $\theta$ term and the phase of the determinant of the quark mass matrix: 
 $\theta \to \theta+ \arg\det{\cal M}_{\text{q}}$.
  Even if canceled 
 by the Peccei-Quinn mechanism,\cite{Peccei:1977hh,Peccei:1977ur} 
 small contributions might reappear generated by BSM physics and by possible `Peccei-Quinn  breaking' terms from 
 Planck-scale physics.}
  of the order of $10^{-10}$ which would  still be compatible with the empirical 
 bound for the neutron EDM\cite{Baker:2006ts} 
 cannot help  in generating a sufficient baryon--antibaryon asymmetry:
 there would be a mismatch of the scales relevant for an electroweak (or even higher)  transition  on the one hand 
 and the $\sim$ 1-2 GeV regime where QCD becomes 
 sufficiently non-perturbative such that
 instanton effects are not suppressed any longer on the other hand, see e.g. Ref.~\refcite{Kuzmin}; 
 \item at vanishing chemical potential the SM, which as  a relativistic quantum field theory is of course $CPT$ invariant,
shows only a rapid cross-over instead of a phase transition of first order. 
\end{romanlist}
Therefore, the  observed matter-antimatter asymmetry
together with the insufficient $CP$ violation of the SM
represents one of the few existing indicators that there might be {\em New Physics}
beyond  the Standard Model (BSM physics) which in turn might imply EDM values for 
subatomic particles that are larger in magnitude than those predicted by the KM-mechanism of the SM. Note that this ``evidence'' for substantial
EDM values (especially for hadrons and nuclei) is at best
circumstantial and by no means compulsory.

 The  current status on 
 the experimental bound of the neutron EDM is $|d_n| < 2.9 \cdot 10^{-26}\,\ecm$ as measured 
 by the Sussex/RAL/ILL group\cite{Baker:2006ts}. It  cuts by two orders of
magnitude into the new physics window (\ref{window}), excluding in this way 
already some simple and minimal variants of the above mentioned
New Physics models, especially some variants of (minimal) SUSY.
The corresponding bound
for the proton, namely $|d_p| < 2.0\cdot 10^{-25}\,\ecm$, is inferred from a theoretical calculation\cite{Dmitriev:2003kb}
applied 
to the input from the 2016 EDM bound for the diamagnetic ${}^{197}$Hg atom, 
$|d_{\text{Hg}}| < 7.4 \cdot 10^{-30}\,\ecm$~\cite{Graner:2016ses}. The same method would predict
for the neutron the bound  $|d_n| < 1.6 \cdot 10^{-26}\,\ecm$ which is even slightly less than the 
Sussex/RAL/ILL limit\cite{Baker:2006ts} but is of course affected by the imponderables in extracting
the relevant nuclear matrix elements of the $^{197}$Hg nucleus.
 The EDM bound on the electron is again inferred indirectly, since
theoretical calculations\cite{PhysRevA.78.010502,PhysRevA.84.052108,Skripnikov:2013} are needed
to deduce it from the corresponding EDM bounds on  paramagnetic atoms, e.g.~${}^{205}$Tl  with
 $|d_{\text{TL}}| < 9.4 \cdot 10^{-25}\,\ecm$ \cite{Regan:2002ta}, or on polar molecules, as 
 e.g.~YbF~\cite{Hudson:2011zz,Kara:2012ay} or the recent  ThO measurement by the ACME group~\cite{Baron:2013eja}. 
 The latter gives the most stringent bound on the electron EDM, 
 $|d_e| < 8.7 \cdot 10^{-29}\,\ecm$.

All the measurements mentioned above have in common that they  only apply  to overall 
charge-neutral states, since the corresponding
particles
can be confined at rest in a trap even in the presence of reversible external electric fields (and a weak holding magnetic field) which are needed to extract the EDM signal.  
In order to trap charged particles (e.g.\ the proton, deuteron or  helion), which would just be accelerated by a constant
electric field, storage rings (see e.g.\ Refs.~\refcite{Farley:2003wt,Semertzidis:2003iq,Pretz:2013us})
or -- as in the case molecular ions --
 traps with rotating electric fields\cite{MolecularIons}  have to be applied.
In fact, as a byproduct of  $(g-2)_\mu$ measurements in storage rings, there already exist a very weak bound on the
EDM of muon,~\cite{Bennett:2008dy}   $|d_\mu|  < 1.8 \cdot 10^{-19}\,\ecm$, as compared  to 
the SM estimate of $\sim 10^{-35}\,\ecm$.

 \section{EDM Sources Beyond the KM Mechanism}
If a nonzero  permanent EDM could eventually be inferred from  some measurement, we would then know that the source behind
 the pertinent $T$ violation (or $CP$ violation if $CPT$ holds) would most likely not be the KM mechanism of the SM (recall
 \eref{window}), but we would still be unable to pin down the very $CP$ violating mechanism:
 it could be genuine New Physics (as SUSY, two-Higgs models, left-right symmetric models) or  just the QCD 
 $\theta$ term, if the relevant  angle were 
in the window $10^{-10} \gtrsim | \bar\theta|\gtrsim 10^{-14}$ (see below).
In the case of genuine New Physics, the scale of the relevant $CP$ violating operator(s) would have to be larger than
the electroweak scale, probably even larger than what is accessible by LHC physics.

However, by matching possible candidate-models
of $CP$ violating physics at these high scale(s) to the coefficients of  SM 
operators of dimension-six\footnote{The solely existing $CP$ violating operator of dimension five is a Majorana mass term which is only relevant for neutrino 
physics.\cite{Weinberg:1979sa}} 
and higher, the machinery of effective field theories (EFTs) and the renormalization group 
can be applied.\cite{Dekens:2013zca} In a repeating chain, the relevant operators, which also mix under this procedure, can  
perturbatively be run down until  a (SM) particle threshold is reached  (subsequently the top quark, Higgs boson, $W^\pm$
and $Z$ bosons, and finally the bottom and charm quarks), where the corresponding particle should be integrated out 
and the coefficients of the operators should be matched to those containing only the remaining active SM
degrees of freedom.
This cascading perturbative  procedure has to stop when the realm of non-perturbative QCD is reached somewhere between 2\,GeV and 1\,GeV, say.
At this chiral scale $\Lambda_\chi$  the pertinent EFT Lagrangian of dimension-six\cite{Dekens:2014jka} can be written as\footnote{In the following, we will concentrate
on the EDM contributions to nucleons and light nuclei. Therefore, the terms involving leptons  and strange quark
contributions are not listed.} 
\begin{eqnarray}
{\cal L}^{\not{T},\not{P}} &=& -\bar\theta \frac{g_s^2}{64\pi^2}\epsilon^{\mu\nu\rho\sigma}G^a_{\mu\nu}G^a_{\rho\sigma}
\nonumber \\
&&  \mbox{}-\frac{i}{2} \sum_{q=u,d} d_q \bar q \gamma_5 \sigma_{\mu\nu}F^{\mu\nu} q
       -\frac{i}{2} \sum_{q=u,d} \tilde d_q \bar q \gamma_5 \half \lambda^a \sigma^{\mu\nu}G^a_{\mu\nu} q \nonumber\\
&&     \mbox{} + \frac{d_\text{W}}{6} f_{abc} \epsilon^{\mu\nu\alpha\beta}
 G^a_{\alpha\beta} G^b_{\mu\rho} G^{c\,\rho}_{\nu} +\sum_{i,j,k,l=u,d} C_{ijkl}^{4\text{q}}{\bar q}_i \Gamma q_j {\bar q}_k \Gamma' q_l  \label{LCPviol}
\end{eqnarray}
where we included, for completeness, the dimension-four QCD $\theta$ term as well. The relevant quantities are
 the quark fields of flavor $q$, the field strength tensors $F_{\mu\nu}$ and $G^a_{\mu\nu}$ of the photon
and the gluon, respectively, the color SU(3) structure constants $f_{abc}$ and Gell-Mann matrices $\lambda^a$. The
various Lorentz structures of the
matrices $\Gamma$ and $\Gamma'$ ensure that the dimension-six four-quark operators, which have net zero flavor,
violate the $P$ and $T$ symmetries.\cite{Dekens:2014jka}. 
The coefficients $d_q$ of the quark EDM terms  and $\tilde d_q$ of the quark chromo-EDM terms scale as
$\sim v_{\text{EW}}/\Lambda^2_{\not{T}}$ while the coefficient $d_{\text{W}}$ of the gluon chromo-EDM term, the so-called Weinberg term,
and the coefficients $C_{ijkl}^{4\text{q}}$ of the  four-quark EDM terms
scale as $\sim 1/\Lambda^2_{\not{T}}$.  $\Lambda_{\not{T}}$ is the scale of the underlying $T$ (or $CP$) 
breaking model and $v_{\text{EW}}$ is the electroweak vacuum expectation value which is a relic of the original coupling
to the Higgs field which had to be inserted to preserve the SM symmetries.\footnote{Thus the coefficients
$d_{u,d}$ and $\tilde d_{u,d}$ effectively scale as $\sim m_{u,d}/\Lambda_{\not{T}}^2$ to ensure that chiral
symmetry is preserved in the limit of vanishing current quark masses $m_{u,d}$. 
The quark and quark chromo EDM operators are therefore counted as dimension-six operators.}
Note that via a chiral $U_A(1)$ transformation    
the first term on the right hand side of \eref{LCPviol} can be rotated into the term $- \bar \theta m_{\text{q}}^\ast \sum_{q} \bar q i\gamma_5 q$
where $m_{\text{q}}^\ast= m_u m_d/ (m_u{+}m_d)$ is the reduced quark mass. In this way, it is evident that
all EDM contributions of the QCD $\theta$ term have to vanish in the chiral limit (as it is also the case for
the quark and quark chromo EDM contributions) and that therefore the
nucleon EDM induced by the {\em strong} $CP$ breaking term has to scale as
\begin{equation}
| d_{\text{N}}^{\bar\theta}| \sim \bar\theta \cdot \frac{m_{\text{q}}^\ast}{\Lambda_{\text{QCD}}} \cdot \frac{e}{2m_{\text{N}}} \sim \bar  \theta \cdot 10^{-16}\ecm\,,
\end{equation}
such that the window (\ref{window}) for physics beyond the KM mechanism of the SM together
with the current bound on the neutron EDM\cite{Baker:2006ts} is compatible with  the 
$
 10^{-10} \gtrsim  | \bar\theta| \gtrsim10^{-14}
$
window for  searches of strong $CP$ breaking.

The $\theta$-term contribution, the Weinberg term and two of three four-quark terms are flavor/isospin symmetric,
while the quark and quark chromo EDMs can be separated into  isospin-conserving and isospin-breaking
combinations, respectively. The third four-quark operator stems from left-right symmetric physics which breaks
isospin and also chiral symmetry at the fundamental level -- for more details, see e.g.\ Ref.~\refcite{Dekens:2014jka}.

\section{EDMs in the Non-Perturbative Realm of QCD}
At and below the chiral scale $\Lambda_\chi$, perturbative methods are not applicable any longer in order to continue
towards the hadronic, nuclear or even atomic scales relevant for the EDM experiments, since
the degrees of freedom change from the quark/gluon ones to hadronic ones. The relevant methods which 
allow  for an estimate of the pertinent uncertainties are lattice QCD and 
chiral effective field theory (i.e.\ chiral perturbation theory and its extension to multi-baryon systems). 

Some progress has been made in lattice
QCD calculations of the EDM of the neutron (and in some cases of proton) when it is induced by the $\theta$ term,
see e.g.\ Refs.~\refcite{Guo:2015tla,Shintani:2015vsx}, but the extra\-polation to physical pion masses still seems
to be problematic for specific lattice methods.\cite{Shintani:2015vsx} First lattice results for the quark EDM scenario 
relating the nucleon EDM to  the tensor charges of the quark flavors are promising\cite{Bhattacharya:2015esa} but of course not sufficient to constrain realistic models
of $CP$ breaking.
Recently,  there have been attempts to work out the quark flavor contributions to the nucleon EDMs also for the case of
quark chromo EDMs.\cite{Bhattacharya:2016oqm} 
Lattice estimates of the nucleon EDM resulting
from the Weinberg term or even from the four-quark terms  are still left for future studies, not to mention
lattice computations of the EDMs of (light) nuclei.

What is the situation from the chiral EFT point of view? The first chiral calculations  of the nucleon EDM induced by
the QCD $\theta$ term were already performed in the late seventies of the last century.\cite{CDVW79} 
The results of this calculation
and more modern ones\cite{Pich:1991fq,ottnad} are that the leading and sub-leading $CP$ violating
pion-loop contributions to the isovector nucleon EDM could be more and more pinned down while the isoscalar contribution turned out to be more suppressed.
In fact, the leading chiral loop diverges, inducing a logarithmic scale dependence and the need for finite
counter terms of the same order as the isovector loop contribution to the neutron and proton EDMs. The number of these terms  
can be constrained to two as shown in Ref.~\refcite{ottnad} and confirmed by 
Ref.~\refcite{Guo12} -- even for the three-flavor case. As there do not exist  any  measurement or   theoretical
methods (apart from lattice studies for the $\theta$-term scenario) to constrain these counter-term coefficients other than naive dimensional analysis arguments (which only refer to the magnitude
but not to the sign), there is not much predictive power of the chiral EFT approach for the total single-nucleon EDMs.
Only when theses calculations are coupled with the lattice ones, which have their own problems as mentioned above,
there can be predictions for the $\bar\theta$-induced nucleon EDMs, see Refs.~\refcite{Guo12,Akan:2014yha} and
Ref.~\refcite{Guo:2015tla}.

\section{EDMs of Light Nuclei}
However, the situation is  quite different  for light nuclei as e.g.\  the deuteron or helion. As already observed in the
mid-eighties of the last century in Ref.~\refcite{Flambaum:1984fb}, the same $CP$-violating pion exchange that causes the divergence
in the loop diagrams appears already at the tree-level order in the 
two-nucleon contributions to the EDMs of (light) nuclei,
such that in this application there is no need for sizable counter-terms  which are always local contact-terms by nature. 
In fact, the first $CP$-violating
$NN$ contact terms appear  only at 
next-to-next-to-leading order relative to the contribution of
the corresponding pion exchange diagram, 
see e.g.\ Ref.~\refcite{Bsaisou:2013}.
 
 \subsection{The hadronic parameters}
Using  chiral symmetry and isospin structure arguments  the following chiral EFT Lagrangian  for the (leading)
$P$- and $T$-violating terms including single-nucleon, purely pionic, pion-nucleon and two-nucleon-contact interactions can be 
postulated:\cite{Mereghetti:2010tp,deVries:2011an,deVries:2012ab,Bsaisou:2013,Dekens:2014jka,Bsaisou:2014zwa,Bsaisou:2014oka}
 \begin{eqnarray}
  \mathcal{L}^{\not{T} \not{P}}_{\text{EFT}} &=& -  d_{\text{n}} \bar N(1-\tau_3) S^\mu N v^\nu F_{\mu\nu}
                                                             -  d_{\text{p}} \bar N(1+\tau_3) S^\mu N v^\nu F_{\mu\nu} \nonumber \\
  && \mbox{}+ (m_{\text{N}} \Delta) \pi_3 \pi^2+ g_0 \bar N \vec \tau \cdot \vec \pi N + g_1 \bar N \pi_3 N    \nonumber \\
                                                     && \mbox{}+        C_1 \bar N N \mathcal{D}_\mu (\bar N S^\mu N) 
                                                + C_2 \bar N \vec \tau N \cdot \mathcal{D}_\mu (\bar N S^\mu \vec \tau N) \nonumber\\
                                                  && \mbox{}   + C_3 \bar N \tau_3 N \mathcal{D}_\mu (\bar N S^\mu N) 
                                                           + C_4\bar N  N  \mathcal{D}_\mu (\bar N \tau_3 S^\mu   N) 
                                                           \nonumber\\
                                                           && \mbox{}+ \text{terms of higher order in the chiral expansion,}
 \label{Lag_had}
 \end{eqnarray}
 where in principle the values of the coefficients of the effective Lagrangian (\ref{Lag_had})  
 characteristically depend on the coefficients
of the Lagrangian (\ref{LCPviol})
  (and might eventually be derived by lattice methods). In this way 
 the models for the underlying physics, which again feed
 with different strength into the coefficients of (\ref{LCPviol}), can in principle be disentangled
if sufficiently enough EDM measurements can be matched to sufficiently enough EDM calculations of
 single- and multi-baryon systems. Currently, this step from the EFT Lagrangian (\ref{LCPviol}) to the chiral EFT Lagrangian
 (\ref{Lag_had}) exists only  in rudimentary form for the $\theta$-term case, allowing
  the determination of the $CP$-violating three-pion coefficient $\Delta$ and, respectively,
 the isospin-conserving and isospin-violating $\pi NN$ coefficients $g_0$ and $g_1$ as function
 of  $\bar\theta$, including uncertainties:\cite{Bsaisou:2014zwa,Bsaisou:2014oka}
 \begin{equation}
  \begin{array}{lclcl}
 \Delta^\theta &=& \frac{\epsilon(1-\epsilon^2)}{16 F_\pi m_{\text{N}}} \frac{M_\pi^4}{M_K^2- M_\pi^2} \bar\theta +\cdots &=&  (-0.37\pm 0.09)\cdot 10^{-3} \bar\theta\,, \\
 g_1^{\theta} &=& 8 c_1 m_{\text {N}}\Delta^\theta + (0.6 {\pm} 1.1) \cdot 10^{-3} \bar\theta &=& (3.4 \pm 1.5)\cdot 10^{-3}\bar\theta\,, \\
 g_0^{\theta} &=&  \bar\theta \frac{ \delta m_{\text{np}}^{\text str}(1-\epsilon^2)}{4 F_\pi \epsilon} &=& (-15.5 \pm 1.9) \cdot
 10^{-3} \bar\theta \,. \end{array} \label{coeff-theta}
 \end{equation}
 The involved quantities are  the pion decay constant, the isospin-averaged masses of the nucleon, pion, and kaon,\cite{pdg:2016} 
 the strong neutron-proton mass splitting $m_{\text{np}}^{\text{str}}$,\cite{Walker-Loud:2014iea, Borsanyi:2014jba} the quark mass ratio $m_u/m_d$,\cite{Aoki:2013ldr} which
 feeds into $\epsilon\equiv \frac{m_u-m_d}{m_u+m_d}$, and the ChPT coefficient $c_1$ (related to the nucleon sigma term).\cite{BKM95}  The additional contribution to $g_1^{\theta}$ results from an independent chiral structure\cite{Mereghetti:2010tp} and its stated value was estimated in Ref.~\refcite{Bsaisou:2013}.
 
 However, even in the $\theta$-term scenario, the other coefficients, especially 
 the total neutron and proton EDM values $d_{\text{n}}$ and $d_{\text{p}}$,  but also the isospin-conserving 
 and isospin-breaking $NN$-contact
 coefficients $C_{1,2}$ and $C_{3,4}$, respectively, 
 can only be estimated in magnitude but not in sign, either by naive dimensional
 arguments or by the magnitude of the subleading loop-contributions in the case of the $C_{i}$ 
 coefficients.\cite{Bsaisou:2014zwa,Bsaisou:2014oka} While the coefficients $d_{\text{n}}$ and $d_{\text{p}}$ 
 can in principle be matched to the corresponding lattice QCD calculations which still -- as already mentioned -- are problematic, the
 estimated contributions of the $C_{i}$ terms have to be treated as systematic uncertainties -- even for light nuclei
 and even in the theoretically most simple $\theta$-term scenario.
 
So far there do not exist  similar relations between the other  parameters of the
Lagrangian (\ref{LCPviol}) and the effective chiral Lagrangian (\ref{Lag_had}) in the case of realistic underlying models.  However, for
the case of a minimal left-right symmetric model, because of its inherent isospin-breaking nature,
a cross-relation between the $\Delta$ parameter and $g_0$ and $g_1$ can
 be established:\cite{Dekens:2014jka,Bsaisou:2014zwa,Bsaisou:2014oka}
 \begin{equation}
 \begin{array}{lclcl}
   g_1^{\text{LR}} &= & 8 c_1 m_{\text{N}} \Delta^{\text{LR}} &=& (-7.5 \pm 2.3) \Delta^{\text{LR}}\,, \\
   g_0 ^{\text{LR}} &= &  \frac{\delta m_{\text{np}}^{\text{str}} m_{\text{N}} \Delta^{\text{LR}}}{M_\pi^2} &=& (0.12\pm 0.02)    \Delta^{\text{LR}}\,.
    \label{coeff-LR}
  \end{array}
 \end{equation}
 
  \subsection{The EDMs of deuteron, helion and triton}
The EDMs of the deuteron, helion and triton follow from the multiplication of the coefficients of
the chiral effective Lagrangian (\ref{Lag_had}) (see  (\ref{coeff-theta}) and (\ref{coeff-LR}) for special cases) and 
the nuclear matrix elements calculated in Refs.~\refcite{Bsaisou:2014zwa,Bsaisou:2014oka} and listed in
the Tables \ref{Tab:Deuteron} and \ref{Tab:Helion}, respectively, as
\begin{eqnarray}
  d_{\text{D}} &=& d_{\text{p}} \cdot \NME(d_{\text{p}}^{\text{D}}) + d_{\text{n}} \cdot \NME(d_{\text{n}}^{\text{D}}) 
                             + g_1\cdot \NME(g_1^{\text{D}}) +\Delta\cdot \NME(\Delta f_{g_1}^{\text{D}}) \nonumber \\ 
                         &&    \mbox{} + 
                             \left\{ C_3 \cdot \NME(C_3^{\text{D}}) + C_4\cdot \NME(C_4^{\text{D}}) \right\} \,,
                                  \label{dDeut} \\
    d_{\he}&=&   d_{\text{p}} \cdot \NME(d_{\text{p}}[\he]) + d_{\text{n}} \cdot \NME(d_{\text{n}}[\he])
                          + \Delta \cdot \NME(\Delta[\he]) \nonumber \\
                          &&  + g_0 \cdot  \NME(g_0[\he]) \nonumber 
                            + g_1\cdot \NME(g_1[\he]) +\Delta\cdot \NME(\Delta f_{g_1}[\he]) \nonumber \\ 
                            &&    \mbox{} + 
                             C_1 \cdot \NME(C_1[\he]) + C_2\cdot \NME(C_2[\he]) \nonumber \\  
                         &&    \mbox{} + 
                             \left\{ C_3 \cdot \NME(C_3[\he]) + C_4\cdot \NME(C_4[\he]) \right\}    \,,   \label{d3He}             
\end{eqnarray}
and the analog of \eref{d3He} for the triton case, i.e. $\he \to \trit$. The terms in curly brackets (the $C_3$ and
$C_4$ contributions) are of subleading order because of the additional isospin-breaking and can be neglected in all cases, except for an underlying model which is left-right symmetric.

The first two terms proportional to $d_{\text {p}}$ and $d_{\text{n}}$ are the single-nucleon contributions to the total EDMs.  Since $d_{\text{n}}$ and $d_{\text{p}}$  can independently be determined by separate experiments, these single-nucleon terms can be subtracted from the expressions in (\ref{dDeut}) and (\ref{d3He}) in order to determine
the multi-nucleon contributions of the corresponding EDMs --  just by using experimental input.
The quantities proportional to $\Delta$ are either contributing to genuinely irreducible three-body interactions in the helion and triton cases (which numerically,
however, turn out to be small) or to
 finite and  momentum-transfer-dependent loop corrections (see $\Delta f_{g_1}$)
of the isospin-breaking $CP$-violating pion exchange (proportional to  $g_1$). These corrections are exceptionally large and add up coherently
to the $g_1$ contributions which then factually are governed by  {\em three different} terms with different chiral structures which 
will be difficult to disentangle without chiral EFT methods.

 There appear less terms for the deuteron case since it is only a two-nucleon system (excluding the $\Delta$ three-body term) and since the
 deuteron acts as an isospin filter: the isospin-conserving ($CP$-violating) $g_0$ and $C_{1,2}$  terms are excluded since they only induce a transition to  the 
 Pauli-allowed ${}^1P_1$ intermediate states  which cannot be undone by the coupling of the photon. The isospin-breaking terms, however, are allowed
 since the transition to the ${}^3P_1$ intermediate states can be reversed by the (isovector part of the) photon coupling.

\begin{table}[ht]
\tbl{Contributions  to the deuteron EDM calculated from the N$^2$LO (chiral) $\chi$EFT potential\cite{Epelbaum:2004fk,Epelbaum:2008ga}, the A$v_{18}$
potential\cite{Wiringa:1994wb} and the CD-Bonn potential\cite{Machleidt:2000ge}, respectively, see Refs.~\refcite{Bsaisou:2014zwa,Bsaisou:2014oka} (remember that $e>0$ here).
In addition, unpublished results using the recent  $\chi$EFT potential\cite{Epelbaum:2014efa,Epelbaum:2014sza,Binder:2015mbz}  with terms up to N$^4$LO are presented.
The tabulated values still have to be multiplied
by the corresponding coefficients from the chiral Lagrangian (\ref{Lag_had}) which are listed as `units'.  Each $C_{i}$ has the dimension $[\text{fm}^3]$,
$d_{\text{n}}$ and $d_{\text{p}}$  carry the dimension $[\efm]$, while $g_1$, $g_0$ and $\Delta$ are dimensionless.\label{Tab:Deuteron}}
{\begin{tabular}{@{}cccccc@{}} \toprule
term & N$^2$LO ChPT &  N$^4$LO ChPT&   A$v_{18}$ & CD-Bonn & units \\ \colrule
$\NME(d_{\text{n}}^{\text{D}})$ & $\phantom{-}0.939\pm0.009$ &$\phantom{-}0.936\pm0.008$  & $\phantom{-}0.914$ & $\phantom{-}0.927$ & $d_{\text{n}}$ \\
$\NME(d_{\text{p}}^{\text{D}})$ & $\phantom{-}0.939\pm0.009$ & $\phantom{-}0.936\pm0.008$ & $\phantom{-}0.914$ & $\phantom{-}0.927$ & $d_{\text{p}}$ \\
$\NME(g_1^{\text{D}})$                   & $ \phantom{-}0.183\pm 0.017 $ &$\phantom{-}0.182 \pm 0.002$& $\phantom{-}0.186$ & $\phantom{-}0.186$ & $g_1\,\efm$ \\
 $\NME(\Delta f_{g_1}^{\text{D}})$  & $-0.748\pm0.138$ &  $-0.646\pm 0.023$ &$-0.703$ & $-0.719$ & $\Delta \,\efm$\\
 $\NME(C_3^{\text{D}})$                 & $\phantom{-}0.05 \pm 0.05$                     &$\phantom{-}0.033 \pm 0.001$                     & --    &  --   &  $C_3\, e\,\text{fm}^{-2}$ \\
 $\NME(C_4^{\text{D}})$                 & $-0.05 \pm 0.05$                     &$-0.006\pm 0.007$                     & --    &  --   &  $C_4\, e\,\text{fm}^{-2}$ \\
\end{tabular}}
\end{table}

\begin{table}[htb]
\tbl{Contributions to the helion and trition EDM calculated from the N$^2$LO (chiral) $\chi$EFT potential\cite{Epelbaum:2004fk,Epelbaum:2008ga}, the A$v_{18}$+UIX
potential\cite{Wiringa:1994wb,Pudliner:1997ck} and the CD-Bonn+TM potential\cite{Machleidt:2000ge,Coon:2001pv},  see Refs.~\refcite{Bsaisou:2014zwa,Bsaisou:2014oka}. Further details as in the captions of Table~\ref{Tab:Deuteron}. \label{Tab:Helion}}
{\begin{tabular}{@{}cccccc@{}} \toprule
         term   & nucleus & N$^2$LO ChPT& A$v_{18}$+UIX & CD-Bonn+TM & units\\ \colrule
$\NME(d_{\text{n}})$    & \Hel & $\phantom{-}0.904\pm0.013$ &$\phantom{-}0.875$&$\phantom{-}0.902$& $d_{\text{n}}$\\
              & \Tri&$-0.030\pm0.007$&$-0.051$&$-0.038$& $''$ \\
$\NME(d_{\text{p}})$    & \Hel &$-0.029\pm0.006$&$-0.050$&$-0.037$& $d_{\text{p}}$\\
              & \Tri & $\phantom{-}0.918\pm0.013$ &$\phantom{-}0.902$&$\phantom{-}0.876$& $''$  \\
$\NME(\Delta$) & \Hel&$-0.017 \pm 0.006$&$-0.015$&$-0.019$& $\Delta\, e\, \text{fm}$ \\
              & $\Tri $&$-0.017\pm 0.006 $&$-0.015$&$-0.018$& $''$\\
$\NME(g_0)$    & \Hel&$\phantom{-}0.111\pm0.013$ &$\phantom{-}0.073$&$\phantom{-}0.087$&$ g_0\,\efm$\\
              & \Tri& $-0.108\pm0.013$&$-0.073$&$-0.085$& $''$ \\
$\NME(g_1)$    & \Hel&$\phantom{-}0.142\pm0.019$&$\phantom{-}0.142$&$\phantom{-}0.146$&$ g_1\,  \efm $\\
              & \Tri& $\phantom{-}0.139\pm0.019$&$\phantom{-}0.142$&$\phantom{-}0.144$& $''$ \\
 $\NME(\Delta f_{g_1})$ &$\Hel $ &$-0.608\pm 0.142$&$-0.556$ &$-0.586$ & $\Delta\, \efm $\\
                              &\Tri &$-0.598\pm0.141$&$-0.564$ &$-0.576$ &$''$  \\
       $\NME(C_1)$        & \Hel & $-0.042\pm 0.017$ & $ -0.0014$ & $-0.016$ & $C_1\, e\,\text{fm}^{-2}$ \\    
              & \Tri & $\phantom{-} 0.041\pm 0.016$ & $\phantom{-}0.0014$ 
          & $\phantom{-}0.016$ &  $''$ \\    
       $\NME(C_2)$        & \Hel & $\phantom{-}0.089\pm 0.022$ & $\phantom{-}0.0042$ & $\phantom{-}0.033$ & $C_2\, e\,\text{fm}^{-2}$ \\    
              & \Tri & $-0.087\pm 0.022$ & $-0.0044$ & $-0.032$ &$''$  \\    
       $\NME(C_{3})$        & \Hel/\Tri & $-0.04 \pm 0.03$ & -- &  -- & $C_3\, e\,\text{fm}^{-2}$ \\    
           $\NME(C_4)$        & \Hel/\Tri & $\phantom{-}0.07 \pm 0.03$ & -- &  -- & $C_4\, e\,\text{fm}^{-2}$ \\    
  \botrule
  \end{tabular}}
\end{table}

In contrast to the application of phenomenological nuclear potentials,\cite{Wiringa:1994wb,Pudliner:1997ck,Machleidt:2000ge,Coon:2001pv}
the calculations using chiral potentials\cite{Epelbaum:2004fk,Epelbaum:2008ga} allow for the specification of uncertainties in addition to  central values.\cite{Bsaisou:2014zwa,Bsaisou:2014oka} 
The latter are mostly compatible with
the results of the phenomenological potentials (which agree, where a comparison is possible, with the calculations of other 
groups\cite{deVries:2011an,Song:2012yh,Yamanaka:2015qfa}),  except for the short-range contact terms. These  are very sensitive to
the model-dependent specifics of the short-range repulsion of the phenomenological potentials  -- for more details see Ref.~\refcite{Bsaisou:2014zwa}. We therefore
refrain from showing results of these phenomenological potentials for the isospin-breaking $C_{3,4}$ contributions which are of subleading nature
to start with. Finally, in Table~\ref{Tab:Deuteron} also unpublished results for the recent chiral $NN$ potential\cite{Epelbaum:2014efa,Epelbaum:2014sza,Binder:2015mbz} with terms up to order N$^4$LO are  reported. The values are compatible with the older N$^2$LO calculations but with reduced uncertainties.

\section{Conclusion}
Let us conclude by describing a way to identify  or exclude the QCD $\theta$ term or the left-right symmetric models as
the primary candidate for an underlying $CP$ violation beyond the KM-mechanism of the SM.
This can be achieved solely via measurements of the EDMs of the neutron, proton, deuteron and helion. Note that the 
exclusive measurements of
the single nucleon EDMs will not suffice to achieve this, since any reasonable underlying model will  predict
$d_{\text{p}}$ and $d_{\text{n}}$ to be approximately of the same magnitude and most probably of opposite sign.

However, if experimental information about $d_{\he}$ and  $d_{\text n}$ can be established, then a fit-value of the $\bar\theta$ angle can be extracted
from \eref{d3He} with input from (\ref{coeff-theta}), treating the small contribution of the proton and of the contact terms as systematical uncertainties.
With \eref{dDeut} applied to this result, the nuclear part of the deuteron EDM and, if the proton EDM is measured as well (or calculated by lattice methods),
also the total deuteron EDM can be predicted allowing for a test of  the $\theta$-term scenario (instead or  in addition to numerical lattice tests).  

Alternatively, a measurement of the neutron, proton and deuteron EDM allows to extract the $\bar\theta$ fit-value -- again solely from experiment -- and for
the prediction of the total helion EDM, including uncertainties. This alternative extraction has the  advantage that there are not any systematical 
uncertainties related to the $NN$-contact interactions, since these are `filtered out' in the deuteron case.

The characteristic signal for the QCD $\theta$-term  scenario would be 
\begin{equation}
  d_{\text{D}} -0.94(d_{\text{p}}+d_{\text{n}}) \approx - (d_{\he} - 0.9 d_{\text{n}})  \approx \half (d_{\trit}-0.9 d_{\text{p}})\,.
\end{equation}
The establishment of the last part of this relation is of course rather unlikely, since a triton EDM measurement would be necessary.
At the same time we would have predictions of the coefficients $\Delta^\theta$, $g_0^\theta$ and $g_1^\theta$  (with $g_1^\theta/g_0^\theta \approx -0.2$) 
which can be used as
input for EDM calculations of heavier nuclei.

If the dimension-four  QCD $\theta$ term can be excluded -- this test should always be done as the first one -- then the next simplest step is to test
the left-right scenario which also has a telling signal.
The above described measurements, either the route of $d_{\he}$ and $d_{\text{n}}$ or the route of
$d_{\text{D}}$, $d_{\text{n}}$ and $d_{\text{p}}$ allow to extract the $\Delta^{\text{LR}}$ parameter and to predict the other alternative.
The  characteristic signal of the left-right model would be
\begin{equation}
  |d_{\text{n}}| \approx |d_{\text{p}}| \ll |d_{\text{D}}| \qquad \text{and} \qquad d_{\text{D}}   \approx d_{\he}   \approx d_{\trit}\,,
\end{equation}
which is quite distinct from the $\theta$-term scenario. Furthermore, the ratio $-g_1^{\text{LR}}/ g_0^{\text{LR}}\gg 1$  is very different
from its $\theta$-term counterpart.

If both models can be excluded, then the measured values  of $d_{\text{n}}$, $d_{\text{p}}$ and $d_{\text{D}}$ still allow to
extract an effective coefficient $g_1$ which includes the $\Delta f_{g_1}$ modification.
Using this as an input, a further measurement of $d_{\he}$ would then allow to isolate the value of the coefficient $g_0$.
The ratio $g_1/g_0$ of these values  should be rather different from those predicted in the  $\theta$-term  and in 
the left-right symmetric scenarios, as otherwise one of these case could not be excluded any longer.
The extracted $g_1$ and $g_0$ values can be used to predict EDMs for other nuclei, namely light nuclei as the triton or
heavier ones as measured in the case of diamagnetic atoms if the calculation of the nuclear matrix elements of
these heavy nuclei can eventually
be done with less than 50\% uncertainty, say. More details can be found in Refs.~\refcite{Dekens:2014jka,Bsaisou:2014zwa,Bsaisou:2014oka}.

 \section*{Acknowledgements}

We would like to thank Tom Kuo and Ismail Zahed for the invitation to write this paper and
express our gratitude to 
our collaborators  and colleagues 
Christoph Hanhart, Ulf-G.\ Mei{\ss}ner,  
Jordy de Vries,
Evgeny Epelbaum, Kolya Nikolaev, Werner
Bernreuther,  and Wouter Dekens for sharing their insights into the topics presented here.
A.W. would like to thank the Kavli Institute for Theoretical Physics for hospitality and Barry Holstein and J\"org
Pretz for comments on an earlier version of the manuscript.
This research was supported  by the DFG (Grant No.~DFG/TR-110), the NSFC (Grant No.~11261130311)
through funds provided to the Sino-German CRC 110 ÔÔSymmetries and the Emergence of Structure in
QCDÕÕ, and in part by the National Science Foundation under Grant No.~NSF PHY-1125915.
The resources of  the J\"ulich Supercomputing Center at the Forschungszentrum  J\"ulich in Germany, 
namely the supercomputers JUROPA, JURECA and JUQUEEN, have
been instrumental in the computations reported here.
\bibliographystyle{h-physrev5}   
                                                   
\bibliography{bib_wirzba_short} 
\end{document}